\documentclass[twocolumn,preprintnumbers,amsmath,amssymb,prl]{revtex4}
\usepackage{graphicx}

\begin{document}

\preprint{Appl. Phys. Lett. 91, 163111 (2007)}

\title{Nonequilibrium electron charging in carbon-nanotube-based molecular bridges}
\author{I. Deretzis}
\email{ioannis.deretzis@imm.cnr.it}
\author{A. La Magna}
\email{antonino.lamagna@imm.cnr.it}

\affiliation{Istituto per la Microelettronica e Microsistemi (CNR-IMM)\\
Stradale Primosole 50, Catania 95121, Italy}

\begin{abstract}
We evidence the importance of electron charging under nonequilibrium conditions for carbon-nanotube-based molecular bridges, using a self-consistent Green's function method with an extended H\"{u}ckel Hamiltonian and a three-dimensional Poisson solver. Our analysis demonstrates that such feature is highly dependent on the chirality of the carbon nanotube as well as on the type of the contact metal, conditioning in a nongeneralized way the system's conduction mechanism. Based on its impact on transport, we argue that self-consistency is essential for the current-voltage calculations of semiconducting nanotubes, whereas less significant in the case of metallic ones.
\end{abstract}

\maketitle

The fabrication of carbon nanotube (CNT) field-effect transistors\cite{Mart98} has seen a constantly increasing trend in the last years, due to the advancements in the characterization and manipulation techniques, as well as the industry's ongoing request for prototype devices that could allow for a shift from the actual silicon-based to that of carbon nanoscale electronics. Alongside, the need to theoretically interpret the behavior of such components has attracted a large amount of research, leading to an optimization of the production process in the laboratory. It is therefore well-known by now that equilibrium electron charging effects can play an important role in the operation of CNT-based molecular bridges, provoking Schottky-type potential barriers of non-negligible dimensions\cite{Chen05, Xue03} in the interface area between the carbon nanotube and the contacts. Electron charging however is also present under non-equilibrium conditions\cite{Zahi05} due to both spatially and energetically anisotropic interactions between the source and drain electrodes and the CNT's local density of states at energies close to the electrochemical potentials of the contacts. This type of phenomenon, nonetheless significant for the quantum transport mechanism of systems within the nanoscale, has received a minor attention in the carbon nanotube context.

In order to examine charging phenomena between CNTs and real metallic leads, approaches that go beyond single-electron tight binding models need to be adopted. In this sense, \textit{ab initio} models have established accurate system descriptions\cite{Neme06,Pala07,Tayl01}, bearing though a non-negligible computational load. Alternatively, semiempirical extended H\"{u}ckel approximations have demonstrated various merits when employed for the extraction of CNT transport attributes (e.g. secondary energy gaps in zigzag metallic nanotubes\cite{Kien06}, feasible narrow diameter CNT study\cite{Dere06} etc.), while being capable of a reliable description of molecular conduction under non-equilibrium\cite{Zahi05} in a more affordable way. Here we present self-consistent quantum transport calculations for device structures based on finite semiconducting and metallic CNTs. We couple the non-equilibrium Green's function formalism\cite{Datt95} (based on an extended H\"{u}ckel Hamiltonian) with a full 3D Poisson solver for a realistic representation of the device and contacts' chemistry and the system's electrostatics. The goal of this study is to evidence the role of electron charging under \textit{nonequilibrium} conditions in the presence of different contact metals (Au, Pt and Al), and focus on the variations of the system's transmission probability when tuning the terminal potentials, which represent the external parameters of our model.

Our approach is based on the single particle retarded Green's function matrix $G = [ES - H - \Sigma_L - \Sigma_R]^{-1}$, where $E$ is the scalar energy, $H$ is the `device' Hamiltonian matrix in an appropriate basis set, $S$ is the overlap matrix in that basis set and $\Sigma_{L,R}$ is the self energy, which includes the effect of scattering due to the left $(L)$ and right $(R)$ contacts. A Landauer-type expression can be used for the current calculation in case of coherent transport:

\begin{equation} \label{curr}
I=\frac{2e}{h}\int_{-\infty}^{+\infty}dET(E)[f(E,\mu_L)-f(E,\mu_R)],
\end{equation}

where $T(E)=Tr[{\Gamma_L}G{\Gamma_R}G^\dagger]$ is the transmission as a function of energy, $\Gamma_{L,R}=i[\Sigma_{L,R}-\Sigma_{L,R}^\dagger]$ and $f(E,\mu_{L,R})$ represents the Fermi-Dirac distribution of electrons in the contact at chemical potential $\mu_{L,R}$ (T=$300K$ in this work). For the description of both device and contacts we use an Extended H\"{u}ckel semiempirical Hamiltonian calculated in a non-orthogonal basis set of Slater-type orbital functions\cite{Dere06}. Charging effects can be introduced in the formalism with the inclusion of a self-consistent potential $U_{SC}(\Delta\rho)$ that is added in the bare device's Hamilonian  $H_0$ ($H= H_0 + U_{SC}(\Delta\rho)$).  The $U_{SC}(\Delta\rho)$ term can then be determined by the approach of Zahid \textit{et al}\cite{Zahi05}:

\begin{equation}
U_{SC}(\Delta\rho)=U_{Laplace}+U_{Poisson}(\Delta\rho)+U_{Image}(\Delta\rho),
\end{equation}

where $\Delta\rho$ represents the change in the charge density between the nonequilibrium and the equilibrium conditions ($\Delta\rho=\rho-\rho_{eq}$), and $\rho$ is given by the expression below:

\begin{equation}\label{dens}
\rho=\frac{1}{2\pi}\int_{-\infty}^{+\infty}dE[f(E,\mu_L)G\Gamma_{L}G^\dagger+f(E,\mu_R)G\Gamma_{R}G^\dagger]
\end{equation} 

The calculation of the Poisson term can derive in the framework of the complete neglect of differential orbital theory, using only the Hartree potential for the Coulomb interaction\cite{Zahi05}.  Laplace and Image expressions are determined numerically by solving the $\nabla^{2}U=0$ equation with a finite element method in real space, using the appropriate boundary conditions\cite{Zahi05}. All three potential components are evaluated on the atomic sites of the CNT. Finally, equation \ref{dens} results computationally demanding and therefore contour integration techniques in the complex energy plane\cite{Brand02} for energies smaller than $min(\mu_L,\mu_R)$, as well as Gaussian quadrature formula implementations have been introduced in the model for optimization purposes.

\begin{figure}
	\centering
		\includegraphics[width=\columnwidth]{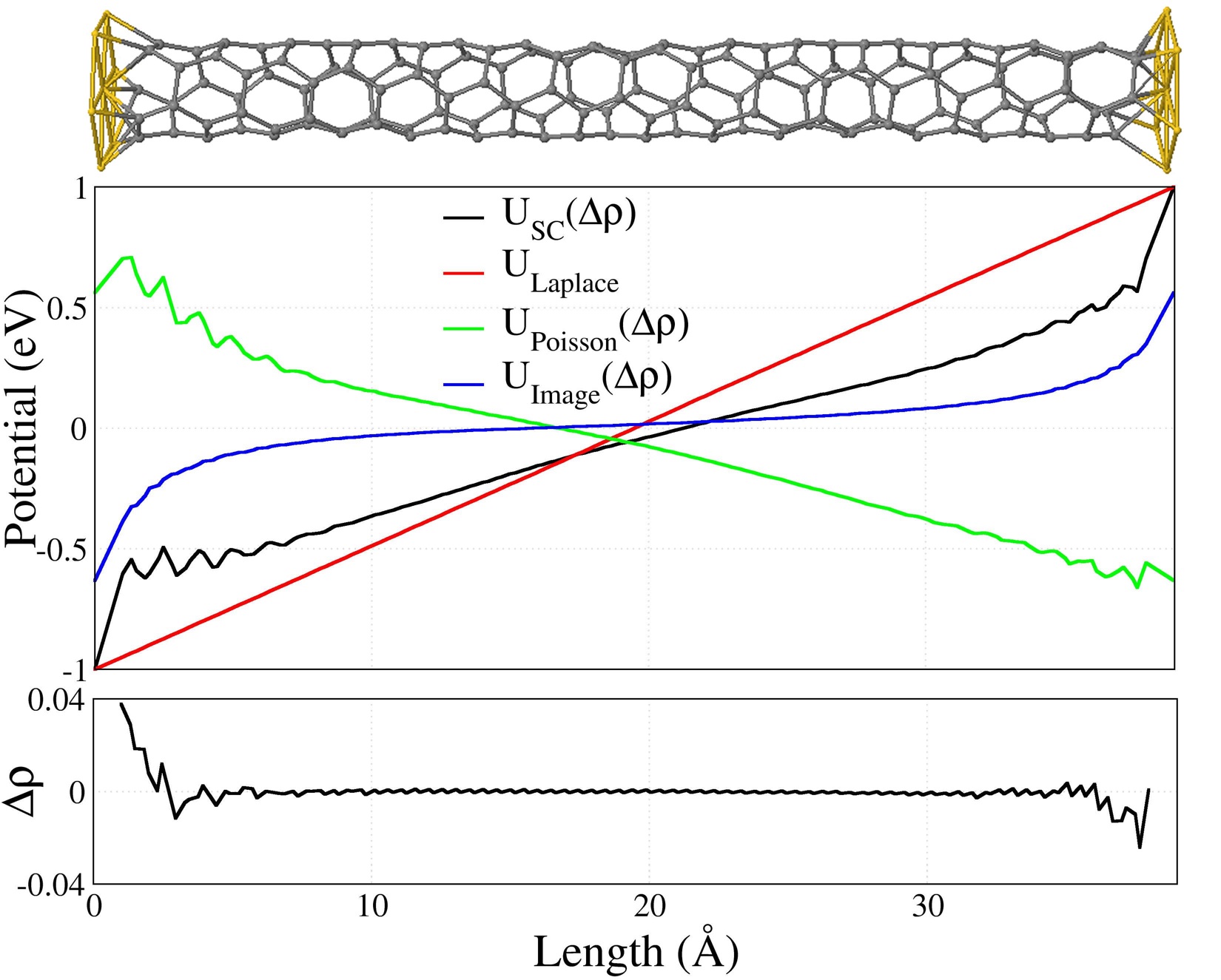}
	\caption{Geometry, potential profile and change of the electronic density $\Delta\rho$ for a 2-unit cell (3,2) CNT with Au(111) metallic contacts, when a $2V$ source-drain bias is applied.}
	\label{fig:figure1}
\end{figure}

We have considered molecular bridges based on defect-free single-wall carbon nanotubes and restricted our calculations in the ballistic regime, since the maximum length of the device tubes is not greater than a few nanometers, less than mean free path measurements\cite{Park04} even for high biases. As the focus of our investigation lies in the nonequilibrium regime, we have separated the nonequilibrium transport features from the electrostatic effects of equilibrium (e.g. Schottky barriers) by imposing the alignment between the metal's work function and the charge neutrality level of the CNT for both equilibrium and nonequilibrium electronic density calculations. This condition can be practically realized shifting the CNT energy bands with respect to the metallic ones, similarly to using a gate electrode to control a CNT field-effect transistor. We have, moreover, set the distance of the source and drain electrodes from the CNT ends to 1\AA{} in order to ensure a device functionality in the self-consistent field regime and avoid undesirable weak coupling effects induced by the finite size\cite{Dere06}, which exceed the current study's objectives.

\begin{figure}
	\centering
		\includegraphics[width=\columnwidth]{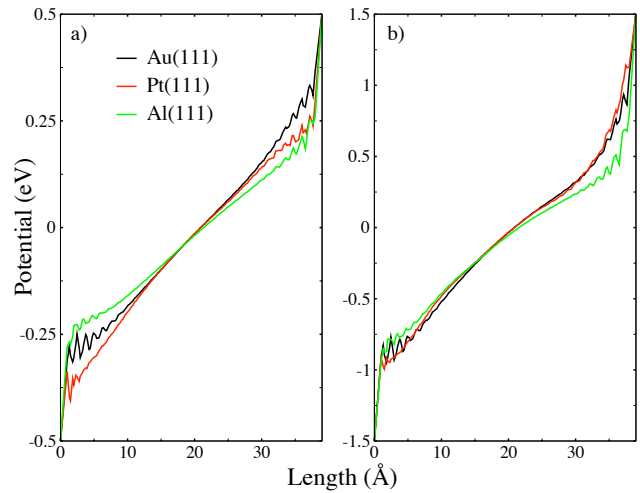}
	\caption{Potential profile of a (3,2) CNT when a 1V $(a)$ and 3V $(b)$ source-drain bias is applied, for Au(111), Al(111) and Pt(111) metallic contacts.}
	\label{fig:figure2}
\end{figure}

\begin{figure}
	\centering
		\includegraphics[width=\columnwidth]{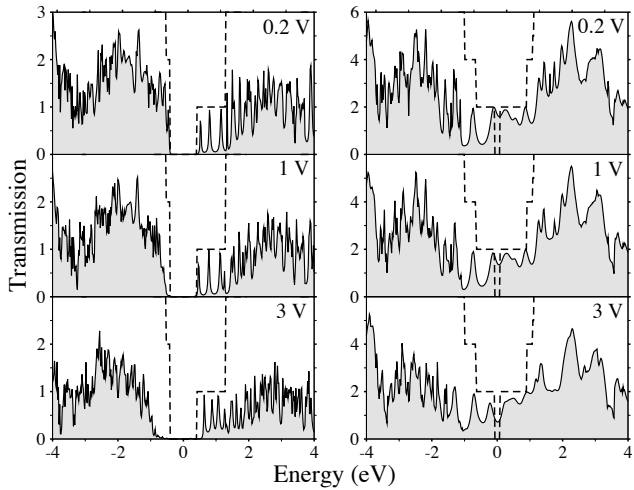}
	\caption{Transmission as a function of energy for a (3,2) 2-unit cell (left column) and a (9,0) 5-unit cell CNT (right column),  for 0.2V, 1V and 3V source-drain biases. The contact metal is Au(111) whereas zero energy refers to the charge neutrality level of each CNT. Interrupted lines represent the transmission probability of respective CNTs in the bulk limit under equilibrium.}
	\label{fig:figure3}
\end{figure}

Figure \ref{fig:figure1} shows the composite potential profile of a (3,2) CNT with Au(111) contacts when a 2V bias is applied, as well as the change of the electronic density $\Delta\rho$ compared to equilibrium conditions. The potential terms that arise due to nonequilibrium charging are the Poisson and Image ones, which consequently compromise the overall term, correcting it with regard to the electrostatic Laplace level. Oscillations of the potential calculated at the atomic sites are due to a higher accumulation of nonuniformly distributed charges in the interface regions that reflect the axially nonsymmetrical interface atom positioning. Our analysis has demonstrated that there are two main factors that shape the aforementioned potential components: a) the chirality of the carbon nanotube and b) the type of the metallic contacts(figure \ref{fig:figure2}). A source-drain bias quantitative dependence has also been observed (see figure \ref{fig:figure2}b). It is interesting to note that the potential drop in the interface area between the nanotube and the contact is also element-dependent and that charging can break the profile's symmetry with respect to the center of the CNT, as can be clearly seen in the case of the $3V$ bias. Figure \ref{fig:figure3} shows how nonequilibrium charging affects the transmission probability of a (3,2) and a (9,0) system in the presence of Au leads. The transmission of the respective bulk systems has also been plotted in order to clarify that nonetheless the small size, the evanescent modes of the system\cite{Pomo04} do not decisively determine the transport features with respect to the long nanotube limit. In both cases and as the bias rises up to $3V$, we can observe a progressive spatial redistribution of the transmission peaks throughout the energy spectrum, which is a direct consequence of the alteration of the system's energy eigenvalues. Although such feature is common for the semiconducting and the metallic CNT, their impact on the conduction mechanism is evidently distinct. Whereas in the metallic tube's case the alteration of the transmission curve does not change the principal transport characteristics, in the semiconducting case we can observe a clear unidirectional motion of the valence band towards smaller energies, which results in a widening of the conduction gap (from about $0.88eV$ with $200mV$ bias to $1.44eV$ with $3V$ bias). Analogous responses are also obtained with the other contact metals, although the positioning of the conduction band is moreover contact dependent.

\begin{figure}
	\centering
		\includegraphics[width=\columnwidth]{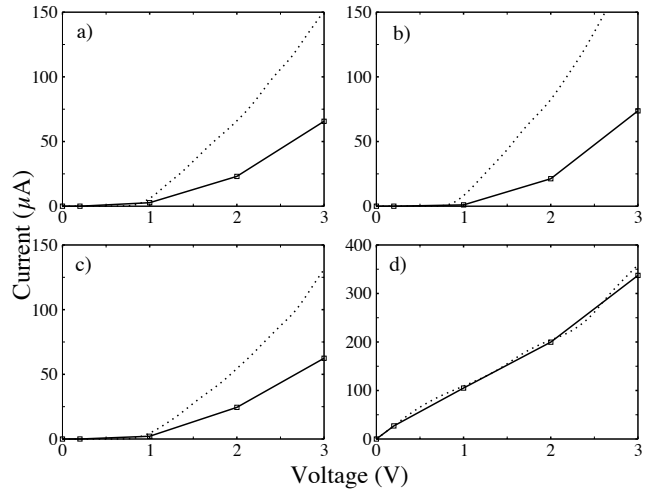}
	\caption{Current-voltage curves for a  (3,2) 2-unit cell in the presence of Au(111) $(a)$, Pt(111) $(b)$ and Al(111) $(c)$ contacts, and for a (9,0) 5-unit cell CNT in the presence of Au(111) contacts $(d)$. Dotted lines represent the respective I-V curves calculated without self-consistency.}
	\label{fig:figure4}
\end{figure}

Charging can be perceived as an interaction between the source and drain electrochemical potentials and the device's local density of states. Since the latter is different in the valence and conduction band zones (see the transmission functions of figure \ref{fig:figure3}), the contacts' electrochemical potentials preferentially correlate with the states of the one band rather than the other, provoking a charge accumulation on the CNT body that can be only captured by a self-consistent approach. This has a practical consequence on the characteristics of the calculated current-voltage curves of the studied CNTs. As we can see in figure \ref{fig:figure4}, in the semiconducting tube's case, a substantial reduction of the overall source-drain current can be noticed with respect to the calculated values without the inclusion of self-consistency (based on the simplified potential model of ref.\cite{Dere06,Zahi03}). Even if the effect can be observed for all contact metals, its impact is variable, being higher in the Pt case rather than in the Au and Al ones. On the contrary, in the metallic tube's case, charging has a minimal influence on the overall conduction mechanism, making possible valid transport calculations without the need to include self-consistency. Finally, early calculations (not shown) for semiconducting and metallic CNTs of different helicities demonstrate a qualitative correspondence to the aforementioned scheme, although results are quantitatively distinguishable from case to case.

To summarize, we have seen how nonequilibrium charging can influence the conduction mechanism of CNT-based molecular bridges, demonstrating that the latter depends on both the chirality of the CNT as well as on the type of the metallic contact. We have moreover showed that the corrections induced in the calculation of the I-V curve are essential in the case of the semiconducting tube as well as significantly contact-dependent, whereas less significant in the metallic tube's case. The importance of such findings can reflect on the common perception of transparency for metals when used as contacts to carbon nanotubes. Our analysis has demonstrated that this cannot be defined unilaterally by the height of the Schottky barrier formation, if other contact-induced features that have an impact on transport are not evaluated. Moreover, apart from carrier-movement phenomena, quantum-chemical effects related to the coupling between the CNT and the metallic electrode(e.g. blocking of conduction channels\cite{Dere06}) can influence strongly the current-carrying capacity of a carbon nanotube. The extent of all aforementioned characteristics is intrinsic to the CNT type. This means that no generalized conclusions can be drawn regarding the quality of a metallic contact when employed in such context. Rather, a more feasible categorization process could be based on self-consistent results obtained for CNTs of similar electrical and geometrical properties (conducting character, charge neutrality level, diameter).

This work was supported by the Regione Sicilia (POR-Regione Sicilia-Misura 3.15).


\begin{thebibliography}{99}

\bibitem{Mart98} R. Martel, T. Schmidt, H. R. Shea, T. Hertel, and Ph. Avouris, \textit{Appl. Phys. Lett.} \textbf{73} 2447 (1998)
\bibitem{Chen05} Z. Chen, J. Appenzeller, J. Knoch, Y.-M. Lin and Ph. Avouris, \textit{Nano Lett.} \textbf{5(7)}, 1497 (2005)
\bibitem{Xue03} Y. Xue and M. A. Ratner,  \textit{Appl. Phys. Lett.} \textbf{83} 2429 (2003)
\bibitem{Zahi05} F. Zahid, M. Paulsson, E. Polizzi, A. W. Ghosh, L. Siddiqui and S. Datta, \textit{J. Chem. Phys.} \textbf{123} 064707 (2005)
\bibitem{Neme06} N. Nemec, D. Tomanek and G. Cuniberti, \textit{Phys. Rev. Lett.} \textbf{96} 076802 (2006)
\bibitem{Pala07} J. J. Palacios, P. Tarakeshwar and D. M. Kim, arXiv:0705.1328v1
\bibitem{Tayl01} J. Taylor, H. Guo and J. Wang \textit{Phys. Rev. B} \textbf{63} 245607 (2001)
\bibitem{Kien06} D. Kienle, J. I. Cerda and A. W. Ghosh \textit{J. Appl. Phys.} \textbf{100}, 043714 (2006)
\bibitem{Dere06} I. Deretzis and A. La Magna, \textit{Nanotechnology} \textbf{17} 5063-5072 (2006)
\bibitem{Datt95} S. Datta, Electronic Transport in Mesoscopic Systems, H. Ahmed, M. Pepper, A. Broers (Eds.), Cambridge University Press (1995)
\bibitem{Brand02} Mads Brandbyge, Jos\'{e}-Luis Mozos, Pablo Ordej\'{o}n, Jeremy Taylor and Kurt Stokbro, \textit{Phys. Rev. B} \textbf{65}, 165401 (2002)
\bibitem{Park04} J.-Y. Park, S. Rosenblatt, Y. Yaish, V. Sazonova, H. Ustunel, S. Braig, T. A. Arias, P. W. Brouwer and P. L. McEuen \textit{Nano Lett.}  \textbf{4} 517 (2004)
\bibitem{Pomo04} P. Pomorski, C. Roland and H. Guo, \textit{Phys. Rev. B} \textbf{70}, 115408 (2004)
\bibitem{Zahi03} F. Zahid, M. Paulsson and S. Datta, Electrical conduction through molecules: Advanced Semiconductors and Organic Nano-Techniques, ed. H Morkoc, New York: Academic (2003), Vol. 3, p. 1

\end{thebibliography}
\end{document}